\documentclass[aps,pra,reprint,onecolumn,notitlepage,eqsecnum,floatfix,showkeys,nofootinbib]{revtex4-2}

\usepackage{amsmath}
\usepackage{amsbsy}
\usepackage{amsfonts}
\usepackage{bigints}
\usepackage{xcolor}
\usepackage{graphicx}
\usepackage{hyperref}
\usepackage{multirow}
\usepackage{linearb}
\usepackage{relsize}
\usepackage{amsthm}
\usepackage{fancyvrb}
\usepackage{calrsfs}
\DeclareMathAlphabet\mathbfcal{OMS}{cmsy}{b}{n}

\newcommand{\bq}{\begin{eqnarray}}
\newcommand{\eq}{\end{eqnarray}}
\newcommand{\bqn}{\begin{eqnarray*}}
\newcommand{\eqn}{\end{eqnarray*}}
\newcommand{\bqs}{\begin{subequations}}
\newcommand{\eqs}{\end{subequations}}
\newcommand{\bw}{\begin{widetext}}
\newcommand{\ew}{\end{widetext}}
\newcommand{\xx}{{\boldsymbol x}}

\newcommand{\rr}{{\boldsymbol r}}

\newcommand{\qq}{{\boldsymbol q}}

\newcommand{\NNNN}{\boldsymbol{N}}

\newcommand{\pp}{{\boldsymbol p}}

\newcommand{\calm}{{\cal M}}
\newcommand{\calo}{{\cal O}}

\newcommand{\cals}{{\cal S}}

\newcommand{\cald}{{\cal D}}

\newcommand{\calp}{{\cal P}}
\newcommand{\calr}{{\cal R}}

\newcommand{\calh}{{\cal H}}

\newcommand{\calu}{{\cal U}}

\newcommand{\calk}{{\cal K}}

\newcommand{\rsc}{R_{\rm Sc}}
\newcommand{\rin}{R_{\rm in}}
\newcommand{\rmcm}{{\rm cm}}
\newcommand{\rmK}{{\rm K}}
\newcommand{\rmg}{{\rm g}}

\newcommand{\vect}[1]{\vec{\boldsymbol #1}}

\newcommand{\id}{1\!\!1}

\SaveVerb{floor}=FLOOR=
\SaveVerb{int}=INT=
\SaveVerb{nint}=NINT=
\SaveVerb{sgn}=SGN=
\SaveVerb{abs}=ABS=

\begin{document}
\title{Thermodynamic and structural properties of the thin atmosphere of a white dwarf}

\author{Riccardo Fantoni}
\email{riccardo.fantoni@scuola.istruzione.it}
\affiliation{Universit\`a di Trieste, Dipartimento di Fisica, strada
  Costiera 11, 34151 Grignano (Trieste), Italy}

\date{\today}

\begin{abstract}
We determine thermodynamic and structural properties of the Coulomb plasma
(jellium) in the thin atmosphere of a white dwarf using the path integral 
Monte Carlo method within a general relativity description of the star.
\end{abstract}


\maketitle
\tableofcontents
\section{Introduction}
\label{sec:intro}

{\sl White dwarfs} \cite{Shapiro-Teukolsky} are the dense remnants of stars, 
composed mainly of carbon and oxygen 
nuclei, which contain electrons, protons and neutrons and are supported against 
collapse induced by gravity by the electron degeneracy pressure due to Pauli 
exclusion principle. They remain white dwarfs till gravity overcomes electron
degeneracy pressure, crushing electrons into protons to form neutrons 
through $\beta^+$ decay.  

Therefore the simplest model for the interior of a white dwarf is that of
{\sl Jellium}. This is a system of pointwise electrons immersed in a 
uniform neutralizing background. The density of the electron gas is so high
that the Fermi temperature is much higher than the temperature of the
star and the electron gas can be considered degenerate. Chandrasekar 
\cite{Chandrasekhar1931,Chandrasekhar1931a,Chandrasekhar1931b,Fantoni17d}
model of a white dwarf equilibrium structure didn't take into account 
the effects of general relativity. First Tolman \cite{Tolman1939}, and 
later Oppenheimer and Volkoff \cite{Oppenheimer1939} proposed a way to 
treat the hydrodynamic equations for the equilibrium and stability of 
a star within the framework of general relativity.   

In order to make some progress towards a more realistic equation of 
state it is necessary to drop the assumptions of an ideal and perfect 
electron gas. This has been accomplished with various mean field theories. 
Since for a white dwarf the Wigner-Seitz radius is small, $0.0003<r_s<0.01$
the corrections due to the Coulomb interaction will be small. 
The main effect of the electrostatic corrections is to give 
smaller radii and larger central densities compared with Chandrasekhar 
models of the same mass.
As described in \S 2.4 of the book \cite{Shapiro-Teukolsky} a first
electrostatic correction arises because the positive charges are not
uniformly distributed in the gas but are concentrated in individual 
nuclei. A more accurate equation of state is needed for the strongly
correlated plasma at high temperature and high density in the classical
regime or at low temperature and low density in the quantum regime. 
Taking linearly into account charge screening in the electron-ion plasma 
a better approximation can be obtained from the Debye-H\"uckel theory
in the classical regime or the Thomas-Fermi (and Thomas-Fermi-Dirac or 
Hartree-Fock) theory in the quantum regime. 
An even more accurate but non linear theory of screening 
is offered by Vlasov theory in the classical regime or 
Lindhard (also known as random phase approximation) theory in the  
quantum regime \cite{March-Tosi}. Apart from these mean field 
theories, perturbative \cite{Gell-Mann1957} or integral 
equation \cite{Singwi1968} approaches are also available.

More progress towards an exact solution of the model of the 
plasma is otherwise
possible employing numerical computer experiments like path 
integral Monte Carlo \cite{Ceperley1995} 
at finite non zero temperature or diffusion Monte Carlo 
\cite{Ceperley1980,Fantoni13g} at zero temperature.
In order to take into account the effects of general relativity
in the equation of state to be used within the Tolman-Oppenheimer-Volkoff
equilibrium it is then necessary to solve a 
path integral Monte Carlo experiment for an electron or an 
electron-ion plasma within general relativity.

We can then proceed following two distinct routes, either we treat the 
quantum particles as living in a classical (non-quantum) spacetime or
classical (non-quantum) particles living in a quantum spacetime
\cite{Fantoni26d}. In this work we propose to use the former route
as we will {\sl discuss} in Sec. \ref{sec:mdm}. 

This welding of {\sl causality}, inherent in general relativity, and 
{\sl casuality}, inherent in the Monte Carlo method, is expected to 
offer an operative approach to the problem of (quantum) statistical 
gravity of many body liquids 
\cite{Fantoni24f,Fantoni25a,Fantoni25d,Fantoni26d,FantoniCQ,Fantoni26x}.
The melt between the superposition principle of quantum physics and
the non linearity of general relativity is still under construction.

The luminosity of a white dwarf is due to a thin radiating layer on the 
star surface (see chapter 4 of the book \cite{Shapiro-Teukolsky}).
The mechanism of a white dwarf cooling is the primary bridge between
our theoretical description of the star and observational data which 
rely on essentially two aspects: luminosity observations and time scale 
of the observed cooling.

{\sl Purpose} of this work is to determine the thermodynamic and 
structural properties of the electron plasma 
in this thin atmosphere using the path integral Monte Carlo (PIMC) 
method \cite{Ceperley1995} at low non zero temperature. In order to take into 
account the effects of general relativity we will use the $3+1$ foliation
of Arnowitt, Deser, and Misner (ADM) \cite{ADM1962} to 
build a matter density matrix from the stress-energy tensor through a path 
integral where on each imaginary timeslice one selects a constant time 
foil \cite{Fantoni26zz}. This amounts to treat the quantum Coulomb plasma
in the classical Schwarzschild spatial metric in a patch of subatomic 
dimensions at the radius $\calr$ of the star of mass $\calm$. In weak
gravity the determinant $^3g$ of the spatial components of the metric 
tensor is such that
$-\ln\sqrt{^3g}\approx -\ln(r^2|\cos\theta|)-\Phi/c^2$ with $\Phi=GM/r$ 
the Newtonian {\sl gravitational potential}. 
We propose to use a {\sl geometrical potential} 
acting on each electron $w=-\hbar[\ln\sqrt{^3g}]/\tau$ with $\tau$ the
imaginary time step used in the discretization of the path integral. 
On the other hand we will neglect the physical gravitational potential 
of the gravitating electrons that should otherwise be included through 
the introduction of the Tolman-Ehrenfest temperature 
\cite{Tolman1930a,Tolman1930b}. 

The problem of the plasma on the surface of a sphere, has already 
been treated in Refs. \cite{Fantoni18c,Fantoni23a,Fantoni26x}. Here
the two dimensional curved surface of the sphere is substituted 
by the three dimensional spatial Schwarzschild foil of the white dwarf. 

We may assume that the electrons dispersion relation is the one
of non relativistic particles $\epsilon(p)=p^2/2m$. This is reasonable
since the density of the electron-proton plasma
\footnote{
Assuming that the white dwarf is made up of a completely ionized plasma 
then we may write $\rho=\mu_e m_u n$ with $n$ the 
electrons number density, $m_u$ the atomic mass unit, and $\mu_e=A/Z$ 
with $Z$ the atomic number and $A$ the mass 
number, so that for example for a $^4$He or a $^{12}$C white dwarf $\mu_e=2$ 
and for a $^{56}$Fe white dwarf $\mu_e=56/26\approx 2.134$.}
in the thin surface layer is $\rho_l\approx 10^3\rmg/\rmcm^3$ (see 
section 4.1 of the book \cite{Shapiro-Teukolsky}) much lower 
than the central densities, corresponding to $r_s \approx 0.2$. Therefore 
$p_F^2/m^2c^2\approx 0.01$ where $m$ is the electron mass and 
$p_F=h(3 n/4\pi)^{1/3}$ is the polarized electrons Fermi momentum.
The temperature of the surface layer falls in the range 
$10^6\rmK\lesssim T\lesssim 10^7\rmK$.
We will consider a white dwarf with a 
radius $\calr\approx 700 {\rm Km}$ and a mass 
$\calm\approx 2\times 10^{30} {\rm Kg}$, so the Schwarzschild radius
is about $\rsc\approx 3 {\rm Km}$.

We will determine the effect of temperature, quantum statistics, gravity,
and of imposing charge neutrality on the weakly correlated electron 
plasma in the quantum regime. The plasma we are confronting with 
move from the lower left corner of the phase diagram of figure 1 of
Ref. \cite{Brown2013} for the Euclidean jellium. 

The paper is organized as follows: In Sec. \ref{sec:mdm} we discuss the
matter density matrix approach; in Sec. \ref{sec:PIMCF} we define the PIMC
used in the simulations; in Sec. \ref{sec:PIMCC} we specify the 
characteristics of the patch; in Sec. \ref{sec:nr} we present our numerical 
results; Sec. \ref{sec:conc} is for conclusive remarks. 
 
\section{Matter density matrix and the PIMC method}
\label{sec:mdm}

In Ref. \cite{Fantoni26d} we discussed a thermal equivalence principle
that should hold when dealing with a statistical physics theory of a
many body fluid within the framework of general relativity: Einstein field 
equations $G_{\alpha\beta}=(G8\pi/c^4)T_{\alpha\beta}$ suggest that the 
statistical 
properties {\sl of spacetime}, determined by the Einstein-Hilbert action
\cite{Fantoni24f,Fantoni25a,Fantoni25d}, and the statistical properties 
{\sl of matter} moving in it, determined solely by the right hand side of 
Einstein equations, the stress energy tensor, should be equivalent, one 
should be the shadow of the other. In particular when looking
at the right hand side we may assume for the density matrix of the many
(scalar) particle system the following expression
\bq \label{eq:dm}
\hat{\rho}=\exp\left(-\int_{\partial\Omega}
\hat{T}^{\alpha\beta}\,\beta_\alpha\,dS_\beta\right),
\eq
that holds in quantum statistical mechanics within special relativity 
\cite{Becattini2012}. In Eq. (\ref{eq:dm}) we use Einstein convention 
to tacitly assume a sum over repeated indexes. We will assume that
this expression continues to hold also within
general relativity. We will use a hat to denote the corresponding 
operator, $\partial\Omega$ to denote an arbitrary spacelike hypersurface 
bounding the four dimensional region $\Omega$ of the curved spacetime 
occupied by the $N$ particles, and $\beta^\alpha$ a vector such that 
$\sqrt{\beta_\alpha\beta^\alpha}=1/k_BT$ with $k_B$ Boltzmann constant 
and $T$ the absolute temperature. 

Within Arnowitt, Deser, and Misner (ADM) $3+1$ foliation of space and
imaginary time
\cite{Fantoni25d} we can evaluate the integral at the exponent of Eq.
(\ref{eq:dm}) choosing a hypersurface orthogonal to the time component 
so that $dS_0=\prod_{\mu=1}^N\sqrt{^3g(\rr_\mu)}\,d\rr_\mu$ where 
$^3g={\rm det}\{g_{ij}\}>0$ with $g_{ij}$ the metric on the spatial 
components and $\rr_\mu$ the {\sl shifted} spatial positions of the $\mu$
particle, with $d\rr_\mu=d\xx_\mu+\NNNN cdt_\mu$. Here 
$\vect{x}_\mu=(t_\mu,\xx_\mu)$ is the 4--vector event for particle
$\mu$ and $\NNNN$ is the {\sl shift} in the ADM foliation
\bq \nonumber
ds^2=g_{\alpha\beta}dx^\alpha dx^\beta=
L^2c^2dt^2+g_{ij}(dx^i+N^ic\,dt)(dx^j+N^jc\,dt),
\eq
with
\bq \label{eq:Sadm}
||g_{\alpha\beta}||=\left(\begin{array}{cc}
L^2+N^iN_i & N_i\\
N_i & g_{ij}
\end{array}\right),
\eq 
where $L$ is the {\sl lapse} and $N^i$ the $i$th spatial component of 
the shift (see Eq. (4) in \cite{Fantoni25d}). By Birkhoff theorem the 
vacuum solution of Einstein field equations outside 
a mass $\calm$ is the one of Schwarzschild for which the shifts are zero
$L^2=(1-\rsc/r)$ and 
$g_{rr}=(1-\rsc/r)^{-1}, g_{\theta\theta}=r^2,
g_{\varphi\varphi}=r^2\cos^2\theta$ with $\rsc=2G\calm/c^2$ the 
Schwarzschild radius with $\calm$ the mass of the star inside 
the spherical shell, by Birkhoff theorem. So that 
$^3g=(1-\rsc/r)^{-1}r^4\cos^2\theta>0$
with $r\gg\rsc$. Where for a typical white dwarf $\calr/\rsc\sim 10^3$.
We will generally use a Greek index to denote one of the
4 spacetime components and a Roman index to denote one of the 3
spatial components. A bold face symbol denotes its status as a
three dimensional vector.

We may then introduce
\bq
\tilde{g}_{ij}(R)&=&
g_{i_1j_1}(\rr_1)\otimes\ldots
\otimes g_{i_Nj_N}(\rr_N),\\
\tilde{g}(R)&=&\prod_{\mu=1}^N\mbox{}^3g(\rr_\mu),
\eq
where $||\tilde{g}_{ij}||$ is a matrix made of $N$ diagonal blocks
$||g_{i_\mu j_\mu}||$ with $\mu=1,2,\ldots,N$ and 
$R=\{\rr_1,\ldots,\rr_N\}$ the shifted spatial positions of the 
$N$ particles. 

By the definition of the stress-energy tensor
\bq
\int\hat{T}^{00}\,dS_0=\hat{\calh}_N,
\eq
where $\hat{\calh}_N$ is the Hamiltonian operator of the $N$ particles
{\sl without} their gravitational potential energy.
For a fluid invariant under translations and rotations 
\footnote{Here we mean invariant under the relative killing vector or
isometry, i.e. a transformation that leaves the metric tensor unchanged.} 
this is the only stress-energy tensor component entering the density 
matrix in Eq. (\ref{eq:dm}) \cite{Becattini2012}.

We then reach the following expression
\bq \label{eq:dm1}
\hat{\rho}=\exp\left(-\beta_0\hat{\calh}_N\right),
\eq
satisfying to Bloch equation
\bq \label{eq:bloch}
\frac{\partial\hat{\rho}}{\partial\beta_0}&=&-\hat{\calh}_N\hat{\rho},\\
\hat{\rho}(0)&=&\id,
\eq
where $\id$ the identity matrix.
Here we choose $\beta_0\sqrt{\tilde{g}^{00}}=1/k_BT$ or 
$\beta_0(R)=1/k_B\tilde{T}(R)=\sqrt{\tilde{g}_{00}(R)}/k_BT$ with
$\tilde{T}(R)=T/\sqrt{\tilde{g}_{00}(R)}$ the Tolman-Ehrenfest
temperature \cite{Tolman1930a,Tolman1930b}. This procedure is quite
unorthodox since it assumes already a position representation 
for the temperature but as we will see later it correctly 
{\sl includes} the gravitational potential energy. On the 
other hand it neglects the self-gravitation of the particles. 
In order to include them one
should solve a self consistent field problem. Instead of treating the 
spacetime metric as a fixed background, the gas particles themselves 
act as a source of stress-energy that alters the spacetime curvature
\cite{Tolman1939,Oppenheimer1939,Andreasson2011}.

The coordinates, $R$, representation of the density matrix (\ref{eq:dm1})
can then be expanded into a path integral \cite{Ceperley1995}
\bq \label{eq:pip}
\rho(R,R';\beta_0)=e^{-\cals_N/\hbar}
=\int\rho(R,R_1;\tau)\cdots\rho(R_{M-1},R';\tau)\,
\prod_{k=1}^{M-1}d^{3N}R_k,
\eq
where $\cals_N=\int_0^{\hbar\beta_0}H_N\,dt$ is the action, 
$\tau(R)=\hbar\beta_0(R)/M$ is a small imaginary {\sl timestep}, 
and the {\sl primitive approximation} \cite{Ceperley1995} to the high 
temperature, $M\gg 1$, density matrix
\bq \label{eq:dmpa}
\rho(R,R';\tau)\propto
\sqrt{\tilde{g}(R)}
\exp-\frac{1}{\hbar}\left[\sum_{\mu=1}^Ng_{ij}(\rr_\mu)(r^i_\mu-{r'}^i_\mu)
(r^j_\mu-{r'}^j_\mu)/4\lambda\tau(R)+\tau(R)\,V_N(R)\right],
\eq
where $H_N=K_N+V_N$ is the classical Hamiltonian with $V_N$ 
the potential energy due to the $N$ particles mutual interactions and 
we assumed {\sl non relativistic} $k_BT\ll mc^2$ \cite{Fantoni26d} 
particles with a kinetic energy in a local Lorentz frame 
$-\lambda\sum_{\mu=1}^N\nabla_\mu^2$ with 
$\lambda=\hbar^2/2m$. Fully relativistic particles with a dispersion 
relation in a local Lorentz frame, 
$\epsilon(\pp)=\sqrt{(\pp c)^2+(mc^2)^2}$ with
$\pp=-i\hbar\nabla$, are {\sl non-local} even in the Lorentz frame 
since the Taylor expansion of 
$\epsilon$ in $\pp$ has ever increasing powers of $\pp$. In order to 
restore locality is therefore necessary to imagine for each particle  
the presence of its {\sl antiparticle} \cite{Dirac1931} with negative 
energy $\bar{\epsilon}(\pp)=-\sqrt{(\pp c)^2+(mc^2)^2}$. 
In the action in Eq. (\ref{eq:dmpa}) we recover the classical kinetic 
energy $\sum_{\mu=1}^N\frac{1}{2}m(ds_\mu/\tau)^2$ where 
$ds_\mu$ is the proper distance for the dislpacement of particle 
$\mu$ on the apatial foil in a Tolman-Ehrenfest timestep $\tau$. 
Attempts have also been tried in the fully relativistic case like
the use of Fock `fifth parameter' by Feynmann which nonetheless, 
despite its intuitive and creative nature, brings about some new
physical interpretations not yet tested experimentally like the
Hawking black hole evaporation as described in chapter 25 of 
the book \cite{Schulman}. 

In Eq. (\ref{eq:pip}), at each timeslice 
$t_k=k\tau$, for $k=1,2,\ldots,M-1$, we consider the section of 
spacetime at constant imaginary time and construct the 
high temperature density matrix on that spatial foil. 
Eq. (\ref{eq:dmpa}) reduces to the usual expression in Euclidean space
\cite{Ceperley1995}. It has been found that one may need corrections 
due to the curvature of space and to the van Vleck determinant 
\cite{DeWitt1957,Schulman}. The curvature of a spatial foil of the 
Schwarzschild metric is zero. In fact, contracting Einstein field 
equations in vacuum implies a vanishing scalar curvature, the trace 
of Ricci tensor. Then Einstein field equations in vacuum reduce to 
the vanishing of the Ricci tensor. Therefore also the scalar curvature 
of a spatial foil, at constant time, of Schwarzschild metric must 
vanish. On the other hand the van Vleck determinant in the high
temperature limit reduces to 1.

The most general case requires taking care also of the 
self gravitation of the electrons. One then would 
like to use self consistently the metric $g_{ij}$ obtained solving 
Einstein field equations for each particles configuration at the 
given timeslice generated by the Metropolis random walk 
\cite{Metropolis,Kalos-Whitlock,Ceperley1995}. 

In the high temperature regime one recovers the {\sl classical statistical 
mechanics} For $M=1$ and $R\approx R'$
\bq \label{eq:dmcl}
\rho(R,R';\tau)\propto\sqrt{\tilde{g}(R)}
\exp-\frac{1}{\hbar}\left[\tau(R)\,H_N\right],
\eq
where $\tau(R)=\hbar\beta_0(R)$. 

Up to here we treated the particles as distinguishable. On the 
other hand the properties of a quantum many body fluid depend 
on the kind of quantum statistics, either Bose-Einstein or 
Fermi-Dirac, obeyed by its identical particles. Therefore we are 
rather interested in \cite{Ceperley1995,Ceperley1996}
\footnote{One can antisymmetrize respect to the first, the second,
or both the arguments of the distinguishable density matrix. We here choose the
first case.}
\bq \label{eq:dmf}
\rho_{\rm B,F}(R,R';\beta_0)&=&\frac{1}{N!}\sum_\calp{\rm sgn}(\calp)\rho(\calp R,R';\beta_0),\\
{\rm sgn}(\calp)&=&(\pm 1)^{\sum_{\nu=1}^N(\nu-1)C_\nu},
\eq
where $\calp$ is any permutation of the $N$ particles such that 
$\calp R=(\rr_{\calp 1},\rr_{\calp 2},\ldots,\rr_{\calp N})$, 
with sign ${\rm sgn}(\calp)$. For bosons ${\rm sgn}(\calp)=+1$. 
For fermions an even (odd) permutation has 
${\rm sgn}(\calp)=+1 (-1)$ and an even (odd) number of exchanges 
of a pair of particles.  Any permutations can be broken into 
cycles $\calp=\{C_\nu\}$ where $C_\nu$ is the number of cycles of 
length $\nu$ in $\calp$. We may then introduce a deBroglie thermal 
wavelength 
\bq
\Lambda(R)=\sqrt{2\lambda\beta_0(R)},
\eq
dependent on the point in the $3N$ manifold.
When the size, $\Lambda(R)$, of the many body path $R_1,R_2,\ldots R_M$ 
equals the interparticle spacing, $n^{-1/3}$ 
with $n$ the proper particle number density, the permutations 
will have an effect on the measure of diagonal observables.
This determines a degeneracy temperature
\bq
T_D=\frac{2\lambda n^{2/3}}{k_B}.
\eq 
We may then introduce an adimensional parameter $\Theta=T/T_D$
such that for $\Theta\ll 1$ quantum statistics will be very 
important.

This expression correctly reduces to the density matrix of the 
classical, $\Theta\gg 1$, system of non relativistic, $p/mc\ll 1$, 
particles in weak gravity,
$\Phi/c^2\ll 1$ with $\Phi(r)=G\calm/r$ with $\calm$ the mass of the star.
In fact 
$\sqrt{\tilde{g}_{00}(R)}\approx \prod_{\mu=1}^N[1-\Phi(r_\mu)]
\approx 1-\sum_{\mu=1}^N\Phi(r_\mu)$. 
So that 
\bq
\frac{\tau(R)\,H_N}{\hbar}=\frac{1}{k_BTM}\left(\sum_{\mu=1}^N\frac{p_\mu^2}{2m}-
mc^2\sum_{\mu=1}^N\frac{G\calm}{c^2r_\mu}+V_N(R)+\ldots\right),
\eq
where we neglected terms of second order in the small quantities
$p/mc$ and $GM/c^2r$.

We have previously proposed \cite{Fantoni12e} to consider the 
$\sqrt{\tilde{g}(R)}$ factors as an external geometrical potential
\bq
W_N(R)=-\ln\left[\sqrt{\tilde{g}(R)}\right]\hbar/\tau(R)+\ldots,
\eq
where the dots stand for additional corrections due to the space
curvature and the van Vleck determinant \cite{DeWitt1957,Schulman}.
These contributions can more easily be taken into account by a 
change of coordinates which transform the Laplace-Beltrami 
operator into the flat Laplacian, solving Bloch equation 
(\ref{eq:bloch}) in the new coordinates and then reverting to the 
original ones \cite{Bastianelli2017}. 

The classical limit 
(\ref{eq:dmcl}) has been used in various spatial dimensions to describe
the properties of Coulomb systems. These offer some exact analytic 
solutions in one dimension \cite{Lenard1961,Lenard1962}, 
in two dimensions \cite{Salzberg1963}, and at a 
special value of the coupling constant in two dimensions on various surfaces: 
The plane \cite{Jancovici81}, the cylinder \cite{Choquard81}, the sphere 
\cite{Caillol81}, the pseudosphere \cite{Fantoni03a}, the Flamm paraboloid 
\cite{Fantoni08c, Fantoni12b}. 
\footnote{In $d+1$ spacetime with $d\leq 2$ the Weyl conformal tensor 
vanishes and the Riemann tensor is completely determined by its trace, 
the Ricci tensor. Contracting Einstein field equations in vacuum
implies a vanishing scalar curvature, the trace of Ricci tensor. Then
Einstein field equations in vacuum reduce to the vanishing of the Ricci 
tensor and consequently to the vanishing of the whole Riemann tensor.
It is only in $d>2$ that general relativity looks like a theory of 
the gravitational interaction. Nonetheless Flamm paraboloid is an 
interesting surface of non constant curvature where the Coulomb plasma
still admits an exact analytic solution.}

On the other hand, the path integral (\ref{eq:pip}) with the primitive
approximation (\ref{eq:dmpa}) has been treated for example in Refs.
\cite{DeWitt1957,Schulman} and applied in various geometries to various 
physical systems. For the example of the sphere treated with the Monte 
Carlo method \cite{Kalos-Whitlock} see Refs. 
\cite{Fantoni18c,Fantoni23a,Fantoni26x}. 

In order to extract the statistical mechanics properties of a many (scalar)
particle fluid within general relativity, we chose to work on the right
hand side of Einstein field equations, building a density matrix from the
stress-energy tensor through a path integral where on each imaginary 
timeslice one selects a constant time foil of the $3+1$ ADM spacetime 
foliation. 

This Matter Density Matrix (MaDM) should follow as a shadow the equivalent 
Metric Density Matrix (MeDM) constructed from the Einstein-Hilbert
action, which gives a 
statistical mechanics of spacetime \cite{Fantoni25d}. It is our 
impression that among the two equivalent routes to statistical 
physics this last one offers a mathematically much more cumbersome 
approach \cite{KlauderLNP}. Nonetheless the use of the Markov stochastic
processes, inherent to the Metropolis algorithm \cite{Metropolis}, to
measure a structural or thermodynamic observable of the many body fluid
should make no distinction among the two routes. In fact the Metropolis 
algorithm is very simple and powerful; it can be used to sample
essentially any probability distribution function regardless of analytic
complexity in any number of dimensions.

In the rest of this work we will choose the Tolman-Ehrenfest temperature 
equal to the absolute temperature $\tilde{T}(R)=T$. As we discussed this
amounts to neglect the gravitational potential energy for the interaction
of the electrons with the star. We will also redefine the matter 
density matrix of Eq. (\ref{eq:dmpa}) without the geometrical factor
$\sqrt{\tilde{g}(R)}$ and include it as a measure of integration in the
path integral expression of Eq. (\ref{eq:pip}).

From now on we will use natural Planck units 
$c=G=\hbar=k_B=1$ where $c$ is the speed of light, $\hbar$ is the 
reduced Planck constant, $G$ is the gravitational constant, and $k_B$ 
is the Boltzmann constant.

\section{PIMC in a spatial foil of Schwarzschild metric}
\label{sec:PIMCF}

A many body system composed of $N$ {\sl distinguishable} particles of mass 
$m$ with positions in 
$R=(\rr_1,\rr_2,\ldots,\rr_N)=(\{\rr_l\})$ where each position vector
$\rr_i=(r_l^1,r_l^2,r_l^3)=(\{r_l^j\})=(\{r^{j_l}\})$ in $3$ dimensions. 
On a foil of Schwarzschild solution (\ref{eq:Sadm}) with spatial metric 
tensor $g_{ij}(\rr)$,
the geodesic distance between two infinitesimally close points $R$ and $R'$ is
$d\tilde{s}^2(R,R')=\sum_{l=1}^Nds^2(\rr_l,\rr_l')$ where 
$ds^2(\rr,\rr')=g_{ij}(\rr)(\rr-\rr')^i(\rr-\rr')^j$. 

The Laplace-Beltrami operator on the manifold of dimension $3N$ is
\bq
\Delta_R=\tilde{g}^{-1/2}\nabla_i(\tilde{g}^{1/2}
\tilde{g}^{ij}\nabla_j),
\eq
where $\nabla=\nabla_R$, $\tilde{g}^{ij}$ is the inverse of
$\tilde{g}_{ij}$, i.e. 
$\tilde{g}_{ij}\tilde{g}^{jk}=\delta_i{}^k$ 
the Kronecker delta.

The Hamiltonian $\calh$ of the $N$ non relativistic particles will be
\bq
\calh=-\lambda\Delta_R+V(R),
\eq
where $\lambda=1/2m$ and $V(R)$ is the {\sl Coulomb potential} 
energy of the system, that we here assume only a function of the particles 
positions and bounded from below.

The density matrix $\rho$ of the system obeys to Bloch equation 
(\ref{eq:bloch}) where we now drop the hat on the operators and the
subscript $N$. 
The position representation of the density matrix is then obtained
from $\rho(R,R';t)=\langle R|\rho(t)|R'\rangle$ with 
$\langle R|R'\rangle=\delta(R-R')/\sqrt{\tilde{g}(R)}$ where 
$\delta$ is a $3N$ dimensional Dirac delta function.
In the small imaginary time $\tau$ limit 
the position representation of the density matrix is \cite{Fantoni26x}
\bq \label{eq:rhocurved}
\rho(R,R';\tau)\propto
\tilde{g}(R)^{-1/4}\sqrt{\cald(R,R';\tau)}\,
\tilde{g}(R')^{-1/4}
e^{\lambda\tau\calr(R)/6}
e^{-\cals(R,R';\tau)},
\eq
where $\calr=0$ is the scalar curvature of the Schwarzschild spatial foil
\footnote{The factor depending on the curvature of the manifold is due to 
Bryce DeWitt \cite{DeWitt1957}. 
For a space of constant curvature there is clearly no effect, 
as the term due to the curvature just leads to a constant multiplicative 
factor that has no influence on the measure of the various observables.

Contracting Einstein field equations in vacuum implies a vanishing 
scalar curvature, the trace of Ricci tensor. Then Einstein field 
equations in vacuum reduce to the vanishing of the Ricci 
tensor. Therefore also the scalar curvature of a spatial foil,
at constant time of Schwarzschild metric must vanish.},
$\cals$ the action, and $\cald$ the van Vleck's determinant 
\cite{Vleck1928,Schulman} 
\bq \label{eq:htdm}
\cald_{\mu\nu}&=&-\nabla_\mu\nabla'_\nu\cals(R,R';\tau),\\
\det||\cald_{\mu\nu}||&=&\cald(R,R';\tau),
\eq
where $\nabla=\nabla_R$ and $\nabla'=\nabla_{R'}$. This determinant is 
the Jacobian of the transformation from the initial conditions given by
fixing the pair of momentum and coordinate of each particle to the 
boundary conditions given by specifying the pair of initial and final 
coordinates needed in the coordinate representation. The square root of the 
van Vleck determinant factor takes into account the density of paths among
the minimum extremal region for the action (see Chapter 12 of Ref. 
\cite{Schulman}). The two factors $\tilde{g}^{-1/4}$ are needed 
in order to have for the density matrix a bidensity for which the boundary 
condition to Bloch equation is simply a Dirac delta function.
In the small $\tau\to 0$ limit
\footnote{Remember that the metric tensor is covariantly constant.}
$\tilde{g}(R)^{-1/4}\sqrt{D(R,R';\tau)}\,\tilde{g}(R')^{-1/4}\to(1/2\lambda\tau)^N$.

For the {\sl action} $\cals$, the {\sl kinetic-action} $\calk$, and the 
{\sl inter-action} $\calu$ we have 
\footnote{The expression for $\calk$ is the one of Eq. (24.16) of Ref. 
\cite{Schulman} to lowest order in $R-R'$.}
\bq \label{eq:action}
\cals(R,R';\tau)&=&\calk(R,R';\tau)+\calu(R,R';\tau),\\ \label{eq:kineticaction}
\calk(R,R';\tau)&=&\frac{dN}{2}\ln(4\pi\lambda\tau)+
\frac{d\tilde{s}^2(R,R')}{4\lambda\tau}.
\eq
In particular the kinetic-action is responsible for a diffusion of the
random walk with a single particle variance
equal to $\sigma^2_{ij}(\rr)=2\lambda\tau/g_{ij}(\rr)$.
The inter-action is defined as $\calu=\cals-\calk$ and for potential
energies bounded from below one can resort to the Trotter formula 
\cite{Trotter1959} to reach the {\sl primitive approximation} 
\footnote{See Ref. \cite{Ceperley1995} for a numerical analysis of the
accuracy of this approximation and for its possible refinements.}
\bq
\calu(R,R';\tau)=\tau [V(R)+V(R')]/2.
\eq

Given then an observable $\calo$ we can determine its thermal average
at an absolute temperature $T$ from
\bq \label{eq:measureO}
\langle\calo\rangle&=&g_s{\rm tr}\{\rho(\beta)\calo\}/Z_N,\\
Z_N&=&g_s{\rm tr}\{\rho(\beta)\},
\eq
where $\beta=1/T$, ${\rm tr}\{\ldots\}$ is 
the trace over the spatial variables, $Z_N$ the canonical partition function, 
$g_s=2s+1$ is the spin degeneracy, and we assumed the Hamiltonian independent 
from spin. For {\sl identical} bodies the spin-statistics 
theorem of quantum field theory, dictates that in spatial dimension bigger 
than two, particles with integer spin are bosons (obeying Bose-Einstein 
statistics, symmetric wavefunctions), while half-integer spin particles are 
fermions (obeying Fermi-Dirac statistics, antisymmetric wavefunctions). 
We will only consider polarized electrons that are fermions with $g_s=1$.

The position representation of the density matrix at an imaginary 
time $t=\beta$ is obtained through a {\sl path integral}
\bq \nonumber
\rho(R,R';\beta)&=&\langle R|\rho(\beta)|R'\rangle=\\ \label{eq:pi}
&&\int\rho(R_0,R_1;\tau)\prod_{k=1}^{M-1}\left[\rho(R_k,R_{k+1};\tau)\,
\sqrt{\tilde{g}(R_k)}\,dR_k\right]
\delta(R_0-R)\delta(R_M-R')\,dR_0dR_M,
\eq
where we have discretized the imaginary time $\beta$ into $M$ 
{\sl timeslices} with a small {\sl timestep} $\tau=\beta/M$.  
The volume element of integration, $\sqrt{\tilde{g}(R)}\,dR$, can be
treated by introducing $W(R)=-[\ln\sqrt{\tilde{g}(R)}]/\tau$ as an 
effective {\sl geometrical potential}.
The many body path $R(t)$ is consequently discretized in $M$ {\sl beads} 
$R_k=(\{\rr_{l,k}\})=(\{r_{l,k}^j\})$ at each 
timeslice $k=1,2,\ldots,M$. We will also call {\sl link} a pair of contiguous 
beads. Note that in order to measure an observable through Eq. 
(\ref{eq:measureO})
it is necessary to consider closed paths such that $R(t+\beta)=R(t)$, or rings 
on the foil.

For identical electrons, that satisfy to the Fermi-Dirac statistics, one needs 
to antisymmetrize the distinguishable density matrix (\ref{eq:pi})
\cite{Ceperley1996}. We will then use $\rho_F$ of Eq. (\ref{eq:dmf}).

We will model the Coulomb plasma in the white dwarf atmosphere as an 
electron gas of $N$ electrons, i.e. a jellium where initially we 
neglect the uniform neutralizing background
\footnote{
In a jellium Eq. (\ref{eq:Cp}) should be substituted by
\bq
V(R)=\sum_m\phi_b(r_m)+\sum_{m<n}\phi(\rr_m,\rr_n),
\eq
with $\phi_b$ the potential energy due to the uniform proper charge density
$\rho_b=-ne$ of the neutralizing background as described in Section 
\ref{sec:nb}. Since $\phi_b$ is just a shallow harmonic pull at the 
center of the patch we will initially neglect it.}. 
The Coulomb potential energy is then
\bq \label{eq:Cp}
V(R)=\sum_{m<n}\phi(\rr_m,\rr_n),
\eq
with $\phi$ the Coulomb pair potential between a charge $q_m$ at 
$\rr_m$ and a charge $q_n$ at $\rr_n$. The exact analytical expression 
for $\phi$ is given by the Copson-Linet potential \cite{Copson1928,Linet1976,Linet1986}
\footnote{Compare with the one on a Flamm paraboloid \cite{Fantoni08c}.}
\bq \label{eq:copson}
\phi(\rr_1,\rr_2)=\frac{q_1q_2}{r_1r_2}
\left[\frac{(r_1-\calm)(r_2-\calm)-\calm^2\cos\gamma_{12}}
{\sqrt{(r_1-\calm)^2+(r_2-\calm)^2-2(r_1-\calm)(r_2-\calm)\cos\gamma_{12}-\calm^2\sin^2\gamma_{12}}}
+\calm\right],
\eq
where
\bq \label{eq:cg}
\cos\gamma_{12}=\sin\theta_1\sin\theta_2+\cos\theta_1\cos\theta_2
\cos(\varphi_1-\varphi_2),
\eq
and the Linet topological term $q_1q_2\calm/r_1r_2$ is an isotropic, 
long-range, mutual energy correction ensuring that the presence 
of the Schwarzschild singularity doesn't introduce unphysical 
net charge accumulation
\footnote{\label{foot:1}
Note that since the angular distances are extremely small,
for example of the order of $10^{-17}$ in case A of Table \ref{tab:tq},
we can use the following approximation
\bq
\phi(\rr_1,\rr_2)=\frac{q_1q_2}{r_1r_2}
\left[\frac{(r_1-\calm)(r_2-\calm)-\calm^2}
{\sqrt{(r_1-r_2)^2+2(r_1-\calm)(r_2-\calm)(1-\cos\gamma_{12})}}+
\calm\right],
\eq
where we may expand $\cos\gamma-1$ to first order in $\theta_i,\varphi_i$. 
We are allowed to approximate the $cos\gamma_{12}\approx 1$ in the 
numerator since $\calm\theta_0^2/\calr\sim 10^{-36}$. We are also allowed 
to neglect the term $\calm^2\sin^2\gamma_{12}/\calr^2\approx
(\calm\theta_0/\calr)^2\sim 10^{-39}$. Note that on a computer in order
to be able to calculate the difference $r_1-r_2$ between two electrons
of the patch we need quadruple precision since $\calr\sim 10^{40}$ 
and $\delta\sim 10^{23}$ so we need at least $17$ significant digits. 
Since we have wall boundary conditions on $r$ and periodic boundary 
conditions on $\theta,\varphi$, in order to fulfill the minimum image
convention, given $\theta_1,\varphi_1$ we will choose the image of 
$\theta_2,\varphi_2$ which gives the minimum $1-\cos\gamma_{12}$.}. 
In our white dwarf scenario, in Planck units, 
$\calm\sim 10^{38}\ll\calr\sim 10^{40}$ and this term will be 
small. In the flat spacetime $\calm\to 0$ 
\bq
\phi(\rr_1,\rr_2)\to\frac{q_1q_2}{\sqrt{r_1^2+r_2^2-2r_1r_2\cos\gamma_{12}}}.
\eq

\section{PIMC Simulation on subatomic patch of the star atmosphere}
\label{sec:PIMCC}

From exercise 4.1 of the book \cite{Shapiro-Teukolsky} follows that 
the depth of the thin radiating surface layer of a white dwarf is
\bq \label{eq:delta}
\cald\lesssim 10^{-2}\calr,
\eq
where the typical radius of a white dwarf is 
$\calr\approx 10^{-3} R_\odot - 10^{-2} R_\odot$.

\begin{figure}[htbp]
\begin{center}
\includegraphics[width=6cm]{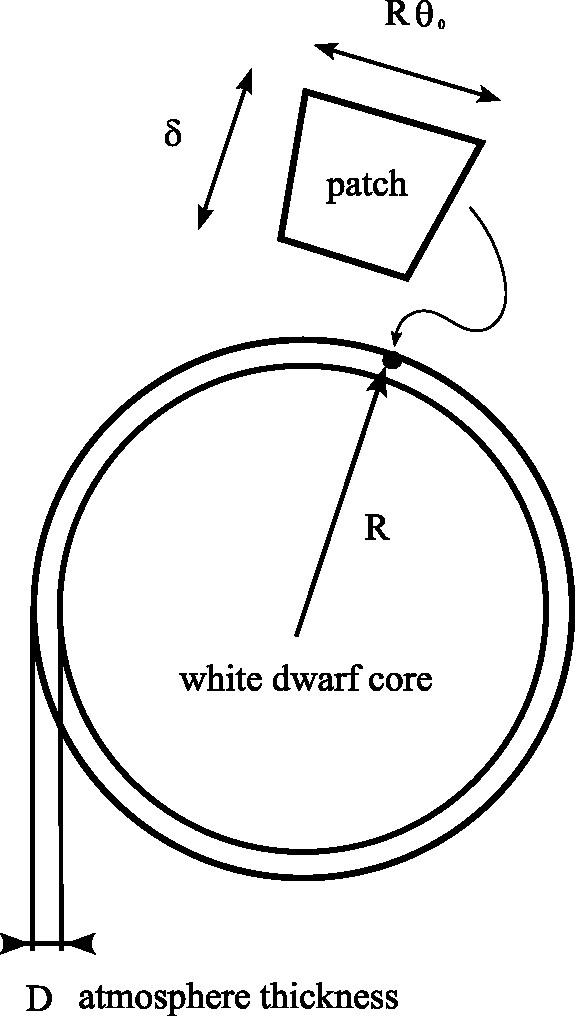}
\end{center}  
\caption{Diagram illustrating the patch used in the simulations.}
\label{fig:patch}
\end{figure}

In order to treat the problem of electrons constrained in a subatomic 
patch of a thin spherical shell of width $\delta\ll\cald$ we imagine an 
infinite repulsive external potential at $r>\calr$ and at $r<\calr-\delta$
In the Monte Carlo we reject any move that brings an electron in these 
regions in oredr to simulate wall boundary conditions on $r$. On the
other hand we will adopt 
periodic boundary conditions on $\theta$ and $\varphi$ with periodicities 
$\theta_0$ and $\varphi_0$ respectively. We will choose $\theta_0=\varphi_0$. 
We want the radial extent of the patch comparable to the angular extent 
so we will choose $\delta\approx\calr\theta_0$. In Fig. \ref{fig:patch}
we show a diagram illustrating the patch.

Of course this is a first brute approximation to a more realistic 
model including an `osmosis' of electrons between the patch 
of the white dwarf atmosphere and the star interior. Using 
periodic boundary conditions on the spherical angles we model a thin
spherical shell of radial width $\delta$. Even if we will not worry 
about finite size errors which occur inevitably due to the choice
of $\theta_0\ll\pi/2$. 

The Cartesian coordinates of a particle on the Schwarzschild spatial foil 
is
\bq
\left\{\begin{array}{l}
x=r\cos\theta\cos\varphi\\
y=r\cos\theta\sin\varphi\\
z=r\sin\theta\\
\end{array}\right.
\eq
where $r^1=r\in[\calr-\delta,\calr]$, $r^2=\theta\in[-\theta_0/2,\theta_0/2]$ 
with $\theta=0$ a Flamm paraboloid equatorial section of the spatial foil, 
$r^3=\varphi\in[-\varphi_0/2,\varphi_0/2]$, and the particle path on it is 
$\qq(t)=(x(t),y(t),z(t))$. The angular extent of the patch 
$\theta_0$ is such that
\bq \label{eq:Omega}
\Omega=\Omega_r
2\sin\left(\frac{\theta_0}{2}\right)\varphi_0\approx
\Omega_r\theta_0^2=N/n,
\eq
where $\Omega$ is the volume of the patch and 
$\Omega_r=\frac{4\pi}{3}\left[\calr^3-(\calr-\delta)^3\right]$.

In natural Planck units $e^2=0.00729735$. Under the experimental 
conditions of case A of Table \ref{tab:tq} we estimate that the 
magnitude of the geometrical potential 
$\sim (1/\beta)\ln(\calr^2)\sim 10^{-24}$ is just one order of 
magnitude larger than the gravitational potential
$Gm\calm/\calr\sim 10^{-25}$ that we will neglect. 
On the other hand, introducing the 
Wigner-Seitz radius $r_s=(4\pi n/3)^{-1/3}/a_0$ with 
$a_0=\hbar^2/me^2$ the Bohr radius, the Copson-Linet pair potential 
$\sim e^2/a_0 r_s\sim 10^{-26}$ is just one order of magnitude smaller 
than the gravitational potential. The kinetic energy is one order of 
magnitude less, $\sim m(r_s/\beta)^2\sim 10^{-27}$.

\subsection{Effect of the neutralizing background}
\label{sec:nb}

The charge neutrality of the jellium in the patch requires
\bq
\rho_b=-\frac{Ne}{\Omega}=-ne,
\eq
Neglecting the angular dependence of the electric field in the patch, 
the Poisson equation for the background potential energy $\phi_b$ is
\bq
\frac{1}{r^2}\frac{d}{dr}\left[r^2\sqrt{1-\rsc/r}\,E\right]=-4\pi\rho_b,
\eq
where $E(r)=-d\phi_b(r)/dr$ is the radial electric field. 
Integrating the Poisson equation from $r=\rin=\calr-\delta/2$ to 
$r$ yields the internal electric field
\bq
E(r)=-\frac{4\pi\rho_b}{3}\frac{r^3-\rin^3}{r^2\sqrt{1-\rsc/r}},
\eq
where we chose $E(\rin)=0$.
Integrating $E(r)$ from $r=\rin$ to $r$ yields the background potential 
profile
\bq \nonumber
\phi_b(r)=-\frac{8\pi e\rho_b}{3}&&\left\{
\frac{1}{2}\left[\sqrt{r^3(r-\rsc)}-\sqrt{\rin^3(\rin-\rsc)}\right]+\right.\\ 
\nonumber
&&\left.\frac{2\rin^3}{\rsc}\left[\sqrt{1-\rsc/\rin}-\sqrt{1-\rsc/r}\right]+\right.\\
\nonumber
&&\left.\frac{3\rsc}{4}\left[\sqrt{r(r-\rsc)}-\sqrt{\rin(\rin-\rsc)}\right]+
\right.\\
&&\left.\frac{3\rsc^2}{4}\left[{\rm atanh}\sqrt{1-\rsc/r}-
{\rm atanh}\sqrt{1-\rsc/\rin}\right]\right\},
\eq
Which reduces to the usual quadratic potential 
$\phi_b(r)\to -8\pi e\rho_b(r-\rin)^2(r+2\rin)/3r$ in 
the flat $\rsc\to 0$ limit, where we grounded the surface $r=\rin$.
The order of magnitude of the background potential for each charge
is $\sim e^2n\delta^2\sim 10^{-26}$.
For simplicity we will neglect the periodic angular images of the 
electrons \cite{Natoli1995}. For case A of Table \ref{tab:tq}
we measured $Ne_K=1.578(7)\times 10^{-24}$ and 
$Ne_V=4.9458(7)\times 10^{-23}$. So the effect of the background is to
increase both the kinetic and potential energy of the plasma. 

\subsection{The black hole limit}
\label{sec:bh}

It is interesting to study the limit of the black hole $\rsc\to\calr$ or
$\calm\to\calr/2$,
keeping all other parameters fixed. By doing so we imagine a star with
a mass increasing up to the point in which its Schwarzschild radius 
coincides with its radius. So that the star becomes a black hole. In
this mathematical limit the geometrical potential diverges to $-\infty$
and, assuming that all the particles are sitting on the horizon at 
$r=\calr$, the Copson pair potential (\ref{eq:copson}) reduces to 
$\phi\to q_1q_2/\calr$, independently of their distance.

In order to get a feeling of the plasma behavior approaching the 
mathematical black hole limit, in our simulations we will increase 
the mass of the star so that, respect to the realistic white dwarf 
scenario, the geometrical potential will become more negative and the 
Copson potential will be diminished.

\section{Numerical results}
\label{sec:nr}

We will choose as reference system (case A of Table \ref{tab:tq}) 
a white dwarf with 
$\calr = R_\odot 10^{-3}$, $\calm = M_\odot$, with an atmosphere with
$n = 3\times 10^{-72}$, $T = 10^6\rmK$.
We will zoom on the patch described in the previous section with 
$r\in [\calr-\delta,\calr]$, $\theta\in [-\theta_0/2,\theta_0/2]$, 
and $\varphi\in [-\theta_0/2,\theta_0/2]$ with the
radial width $\delta\approx\calr\theta_0$. We will use just 
$N=30$ electrons of the atmosphere and we will choose 
$\delta=R_\odot 10^{-20}$ and $\theta_0=10^{-17}$ according to 
(\ref{eq:Omega}).  

We use the Metropolis algorithm \cite{Metropolis,Kalos-Whitlock} to evaluate
the average of Eq. (\ref{eq:measureO}). 

In order to explore ergodically the
positions space $\rr=(r,\theta,\varphi)$ to sample the 
distinguishable density matrix we use the transition 
{\sl displacement} move described in Appendix A of Ref. \cite{Fantoni26x}.

In order to sample the permutation sum of Eq. (\ref{eq:dmf}) needed for 
fermions we use a transition move combination of 2 Brownian 
{\sl bridges} between unlike electrons as described in Appendix B
of Ref. \cite{Fantoni26x}. To construct a single Brownian bridge it 
is essential to map, project, the Schwarzschild spatial foil on a flat 
coordinate system. Our choice is presented in 
Appendix B of Ref. \cite{Fantoni26x}. Perform the Gaussian bridge move in 
the projection flat space and then map it back to the Schwarzschild spatial 
foil. The Metropolis 
algorithm will then allow to sample the high temperature density matrix 
whose kinetic action is not purely Gaussian on the Schwarzschild spatial 
foil, due to the metric tensor appearing in $d\tilde{s}^2$.

We will work in the canonical ensemble with fixed number of particles
$N$, electron number density $n=-\rho_b/e$, and 
absolute temperature $T=1/k_B\beta$. The jellium
{\sl degeneracy parameter} is $\Theta=T/T_D$ where the
degeneracy temperature $T_D=n^{2/3}\hbar^2/mk_B$
\footnote{This is the temperature at which the size of a single 
electron path, in absence of interaction, i.e. the thermal 
wavelength $(2\lambda\beta)^{1/2}$, equals the separation between 
single particle paths, i.e. roughly $n^{-1/3}$. Below this
temperature it is possible for single particle paths to link up
exchanging their end points.}. For temperatures
higher than $T_D$, $\Theta\gg 1$, quantum effects are not very important.
The {\sl Coulomb coupling constant} is $\Gamma=\beta e^2/a_0r_s$. 
At weak coupling, $\Gamma\ll 1$. The plasma becomes weakly correlated
at high $r_s$ in the classical regime, $\Theta\gg 1$, and at
low $r_s$ in the quantum regime, $\Theta\ll 1$, since the kinetic 
energy scales as $1/r_s^2$ and the potential energy as $1/r_s$.

In Table \ref{tab:tq} we report the cases studied. We keep the 
electrons number density $n=3\times 10^{26} \rmcm^{-3}$, the 
number of electrons $N=30$, and the patch geometry fixed and we 
vary the temperature. 
As we can see from the values of $r_s$, $\Gamma$, and $\Theta$ for 
cases A, B, C the atmosphere is expected to be in the quantum 
regime and weakly correlated. Case M is like case A but with 
$\calm=0$ so that the geometrical effect of
gravity is canceled. Case BH is like case A but with 
$\calm=10^{40}$ so that the Schwarzschild radius comes 
closer to the plasma by $\approx 110$ solar masses.

\begin{table*}[htbp]
\protect\caption{Cases treated in our simulations:
  $\calr$ the white dwarf radius, $\delta$ the patch radial width, 
  $\calm$ the white dwarf mass,  
  $N$ the number of particles, $r_s$ the Wigner-Seitz radius,
  $\beta$ the inverse temperature. The adimensional quantities 
  $\Gamma$ and $\Theta$ were introduced in the main text. The 
  results for the thermodynamic quantities, $e_K=\langle K\rangle/N$ 
  from the thermodynamic estimator 
  $K(R,R';\tau)=\tilde{g}_{\mu\nu}(R)(R-R')^\mu(R-R')^\nu/4\lambda\tau^2$
  as explained
  in Ref. \cite{Ceperley1995}, $e_V=\langle V\rangle/N$ the Coulomb 
  potential energy per particle, and $e_W=\langle W\rangle/N$ the geometrical 
  potential energy per particle, are given for the simulation
  without the sum over the permutations in Eq. (\ref{eq:dmf}) and with the
  permutations sum in the bottom table. In the RPIMC case we also report 
  the average number of two particles exchanges $<N$. We neglected the
  presence of the neutralizing background of Section \ref{sec:nb}.
  We chose natural Planck units with $c=G=\hbar=k_B=1$.  
  The simulation lasted not less than $10^5$ Monte Carlo steps
  where one step is made of a displace move of all the beads of a single
  particle path as described in Appendix A of Ref. \cite{Fantoni26x} and 
  a bridge move as described in Appendix B of Ref. \cite{Fantoni26x}.}  
\vspace{1cm}
\label{tab:tq}
\begin{ruledtabular}
\begin{tabular}{|c|lllllllllll|}
\hline
case & $M$ & $N$ & $\calr$ & $\delta$ & $\theta_0=\varphi_0$ & $\calm$ & 
$\beta$ & $a_0r_s$ & $r_s$ & $\Gamma$ & $\Theta$ \\ 
\hline
A & $30$  & $30$  & $R_\odot 10^{-3}$ & $R_\odot 10^{-20}$ & $10^{-17}$ & 
$M_\odot$ & $1.4\times 10^{26}$ & $2\times 10^{23}$ & $0.061$ & $5.17$ & 
$0.03$ \\
B & $300$ & $30$  & $R_\odot 10^{-3}$ & $R_\odot 10^{-20}$ & $10^{-17}$ & 
$M_\odot$ & $1.4\times 10^{27}$ & $2\times 10^{23}$ & $0.061$ & $51.7$ & 
$0.003$ \\
C & $3$ & $30$    & $R_\odot 10^{-3}$ & $R_\odot 10^{-20}$ & $10^{-17}$ & 
$M_\odot$ & $1.4\times 10^{25}$ & $2\times 10^{23}$ & $0.061$ & $0.517$ & 
$0.3$ \\
M  & $30$ & $30$  & $R_\odot 10^{-3}$ & $R_\odot 10^{-20}$ & $10^{-17}$ & 
$0$       & $1.4\times 10^{26}$ & $2\times 10^{23}$ & $0.061$ & $5.17$ & 
$0.03$ \\
BH & $30$ & $30$  & $R_\odot 10^{-3}$ & $R_\odot 10^{-20}$ & $10^{-17}$ & 
$109.4M_\odot$ & $1.4\times 10^{26}$ & $2\times 10^{23}$ & $0.061$ & 
$5.17$ & $0.03$ \\
\hline
\end{tabular}
\end{ruledtabular}
\vspace{1cm}
\begin{ruledtabular}
\begin{tabular}{|c|lll|llll|}
\hline
& \multicolumn{3}{c|}{PIMC no permutations} & \multicolumn{4}{c|}{RPIMC with permutations} \\
\hline
case & $Ne_K(\times 10^{-24})$ & $Ne_V(\times 10^{-23})$ & $-Ne_W(\times 10^{-21})$ & \# swaps & $Ne_K(\times 10^{-24})$ & $Ne_V(\times 10^{-23})$ & $-Ne_W(\times 10^{-21})$\\ 
\hline
A & $1.627(2)$   & $4.6549(1)$   & $1.178895$ & $25.6(2)$   & $1.637(3)$   & $4.6550(2)$   & $1.178895$ \\
B & $1.636(2)$   & $4.6546(2)$   & $1.178895$ & $20.3(6)$   & $1.637(4)$   & $4.6548(3)$   & $1.178895$ \\
C & $1.605(2)$   & $4.6548(1)$    & $1.178895$ & $26.01(3)$ & $1.643(3)$   & $4.6549(2)$   & $1.178895$ \\
M  & $-$   & $-$   & $-$  & $26.0(1)$ & $1.627(2)$   & $4.6548(2)$  & $1.178881$ \\
BH & $-$   & $-$   & $-$  & $26.1(1)$ & $2.104(4)$   & $4.2502(2)$  & $1.180852$ \\
\hline
\end{tabular}
\end{ruledtabular}
\end{table*}

Apart from thermodynamic properties like the {\sl kinetic energy} per 
particle, ${\rm const}-e_K>0$, the {\sl Coulomb potential energy} per 
particle, $e_V$, 
\footnote{The estimators for these observables are carefully described in 
Ref. \cite{Ceperley1995}.}
and the {\sl geometrical potential energy} per particle, $e_W$, 
we will also measure structural properties 
like the {\sl radial distribution function} (RDF), $g(r)=\langle\calo\rangle$. 
For it we used the following histogram estimator,
\bq \label{eq:rdf}
O(R;r,a)=\sum_{m\neq n}
\frac{1_{]r-\Delta/2,r+\Delta/2]}(d_{mn})}{Nn_{id}(r)},
\eq
where $\Delta$ is the histogram bin, $1_{]u,v]}(x)=1$ if $x\in]u,v]$
and 0 otherwise, $d_{mn}$ is the Euclidean distance between 2 electrons 
$m,n$
\footnote{Here it is important to expand $\cos\gamma_{mn}-1$ of 
Eq. (\ref{eq:cg}) 
at least to second order in $\theta_i,\varphi_i$ for $i=n,m$ in order 
to have the correct behavior of the RDF at small distances.} 
\bq \label{eq:ed}
d_{mn}=d(\rr_m,\rr_n)=\sqrt{(\qq_n-\qq_m)\cdot(\qq_n-\qq_m)},
\eq
and 
\bq
n_{id}(r)=4\pi n\left[(r+\Delta/2)^3-
(r-\Delta/2)^3\right]/3~,
\eq
is the average number of particles on the spherical shell
$]r-\Delta/2,r+\Delta/2]$ for the ideal gas of 
density $n$. We have that $n^2g(r)$ gives the probability, that
sitting on a particle at $\rr$, one finds another particle at
$\rr^\prime$ with $r=d(\rr,\rr')$. 

Fermions properties cannot be calculated exactly with path integral
Monte Carlo because of the {\sl fermions sign problem}
\cite{Ceperley1991,Ceperley1996}. We then have to resort to an
approximate calculation. The one we chose was the Restricted Path
Integral (RPIMC) approximation 
\cite{Ceperley1991,Ceperley1996,Fantoni18c,Fantoni21i} 
with a ``free fermions restriction''. The trial density matrix used in the
{\sl restriction} is chosen as the one reducing to the ideal density matrix
in the limit of $t\ll 1$ and is given by the following explicit analytic 
expression,
\bq \label{ifdm}
\rho_0(R_k,R_l;t)\propto{\rm det} \left|\left|e^{
-\frac{d^2(\rr_{i,k},\rr_{j,l})}
{4\lambda t}}\right|\right|.
\eq
Note that here we purposely chose the Euclidean distance (\ref{eq:ed}).
We did this just for the sake of simplicity. In 
any case both the Euclidean and the geodesic distance reduce to 
$ds^2(\rr_{i,k},\rr_{j,l})=
g_{\alpha\beta}(\rr_i)(\rr_{i,k}-\rr_{j,l})^\alpha
(\rr_{i,k}-\rr_{j,l})^\beta$ in the $t=\tau|k-l|\ll 1$ limit. 
The implementation of the RPIMC on a curved space has already been 
described in section IV B of Ref. \cite{Fantoni26x}. The permutation
sampling necessary to compute the sum in Eq. (\ref{eq:dmf}) is carried 
out through the composition of two Brownian bridges and a swap of 
two identical particles (we explicitly neglected 3 particles 
permutation cycles). The realization of a Brownian bridge on the 
curved Schwarzschild space used a mapping onto a flat space as 
explained in appendix B of Ref. \cite{Fantoni26x} with the only
care of distinguishing between a radial standard deviation
$\sigma_r=\sqrt{2\lambda\tau}$ and an angular one 
$\sigma_a=\sigma_r/\calr$ for the Gaussian in the flat projected 
space $(r,\theta,\varphi)$.

We start the simulation with random beads of each electron at each 
timeslice within the subatomic patch. In case A of Table
\ref{tab:tq} we simulate the white dwarf at different temperatures: 
$T=10^6 \rmK$, in case B $T=10^5 \rmK$, and in case C
$T=10^7 \rmK$. Case M is like case A but in absence of with 
$\calm\to 0$. And case BH is like case A but with a larger mass of 
the star approaching the black hole limit described in Section \ref{sec:bh}. 
In all cases we keep a fixed timestep 
$\tau=\beta/M=4.7\times 10^{24}$ in Planck units. For each case
we use either the PIMC treating the electrons as Boltzmannons, 
neglecting the permutations sum in Eq. (\ref{eq:dmf}) and the RPIMC
which correctly but approximately treats the electrons as Fermions.  

From the table we see how the potential, Coulomb or geometrical, 
does not change appreciably between cases A,B,C. On the other hand, 
the kinetic energy increases with temperature in the simulations 
without the permutations and has the opposite behavior when RPIMC 
is used. Quantum statistics affects $Ne_K$ by $\sim 1\%$.
The relative difference of 
the geometrical potential of case A and M is $\sim 10^{-5}$ while 
the kinetic energy relative difference is already $\sim 10^{-2}$. 
Approaching the black hole limit the
geometrical potential becomes more negative, the Copson 
potential is lowered, and the kinetic energy diminishes. This 
is shown by comparing cases A and BH of the Table.

\begin{figure}[htbp]
\begin{center}
\includegraphics[width=10cm]{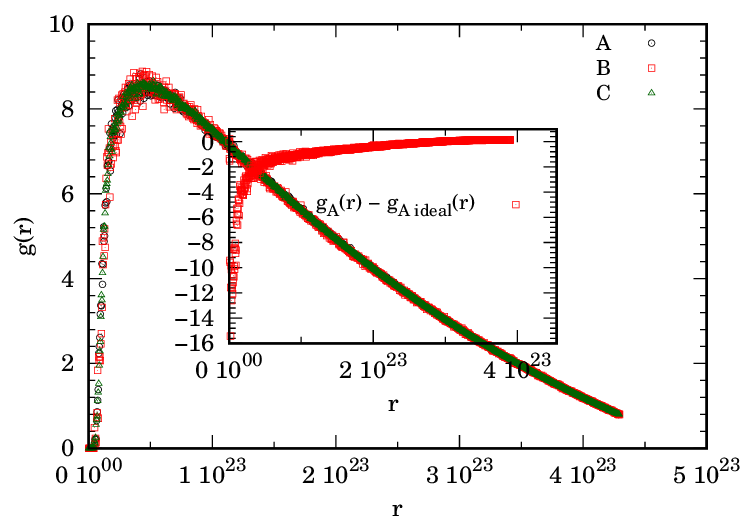}
\end{center}  
\caption{The RDF for cases A,B,C of Table \ref{tab:tq} without 
permutations. We explicitly verified that taking into account the
permutations among the electrons does not change appreciably the
RDF. The RDFs of cases M and BH are also not appreciably different 
from the one of case A shown here. The radial
distance $r$ is in Planck units $1.61623\times 10^{-33}\rmcm$. The 
radial length of the patch is $\delta=4.3\times 10^{23}$. In the inset we show 
the difference between the RDF of case A and the RDF of case A without the 
Copson pair potential between the electrons.}
\label{fig:rdf}
\end{figure}

In Fig. \ref{fig:rdf} we show the RDF for the three cases A,B,C 
of Table \ref{tab:tq} without the electrons permutations. 
The peak of the RDF is at a distance 
$\Delta r\approx a_or_s/4\sim 10^{23}\times 10^{-33}\rmcm 
\sim 10^{-2}$\AA. This means that two electrons are likely to be 
found closer than the Wigner-Seitz radius. From
the figure we see how $g(r\to 0)\to 0$ in agreement with the 
divergence of the Coulomb potential when the two electrons coincide. 
The patch geometry accommodates just the first oscillation of 
the RDF which otherwise would tend to $g(r\to\infty)\to 1$.
The extremely high peak can be explained as due to the 
geometry of the patch. Compare with the RDF of Euclidean 
jellium in a periodic cubic box \cite{Brown2013}. 
We see how the change in temperature does not produce
an appreciable change in the RDF. The same can be said for the
effect of the Fermi statistics or the geometrical effect of gravity.

\begin{figure}[htbp]
\begin{center}
\includegraphics[width=10cm]{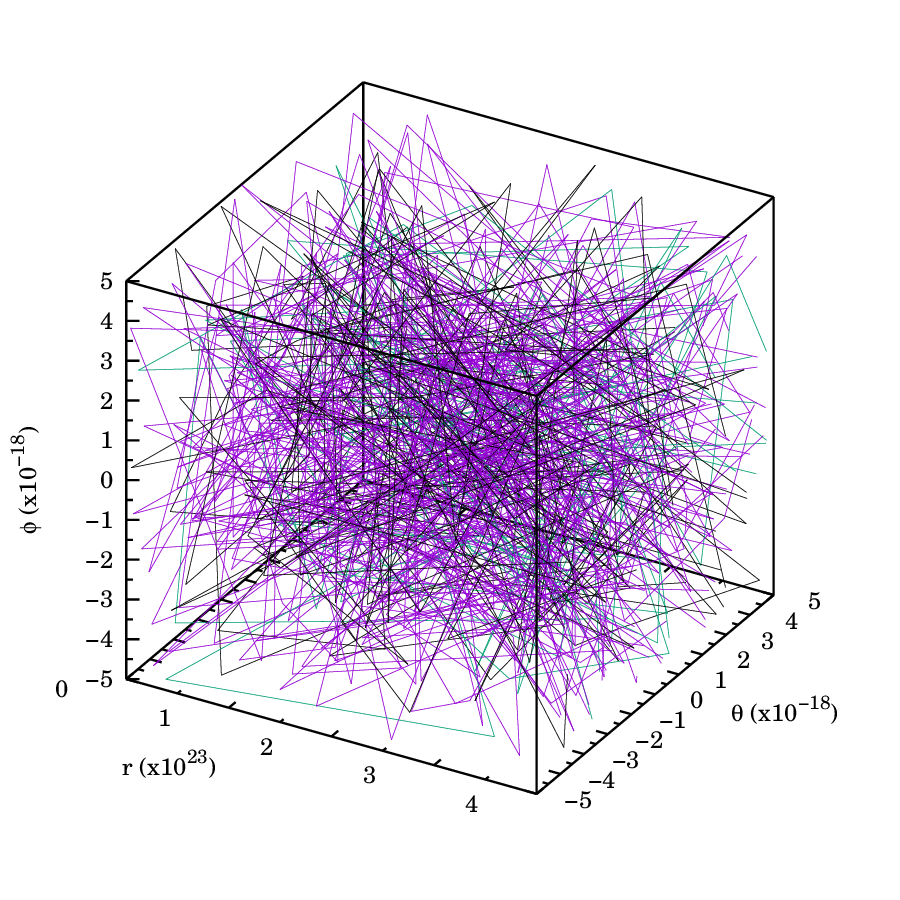}
\end{center}  
\caption{Snapshot of the Coulomb plasma of case A of Table \ref{tab:tq} 
with permutations. 
The radial distance $r$ is in Planck units $1.61623\times 10^{-33}\rmcm$.
The paths of the 30 electrons are painted with different colors for
different permutations cycles.}
\label{fig:splot}
\end{figure}

In Fig. \ref{fig:splot} we show a snapshot of the electrons paths
in the coordinates $(r,\theta,\varphi)$ in the plasma for the patch 
of case A with RPIMC of Table \ref{tab:tq} at thermal equilibrium.

%

\section{Conclusions}
\label{sec:conc}

Still largely unknown is the process of cooling of a white dwarf. We think
that the dead star still possesses some activity in its thin radiating 
atmospheric layer. Which is the source of the luminosity that we are able
to measure. It is therefore important to be able to understand the 
properties, thermodynamic and structural, of the plasma that forms the 
atmosphere. 

In this work we simulated a subatomic patch of this layer. We modeled
the plasma as a jellium where we initially neglect the effect of the 
uniform neutralizing background. For the Coulomb pair potential between the
electrons we adopted the Copson solution for particles living in a spatial 
section of the Schwarzschild metric. We modeled the patch of the dimensions 
of \AA$/20$ using periodic boundary conditions on the spherical angles and
wall boundary conditions on the radial coordinate.
The expected experimental conditions of the plasma suggest that it will be 
weakly interacting and the quantum statistics will play a non negligible role. 
Therefore we used the path integral Monte Carlo method to estimate the quantum 
effects. We found that quantum
statistics affects the third significant figure of the kinetic energy. 
We then used the Schwarzschild solution of Einstein field equations to 
estimate the {\sl geometrical} effect of gravity.
These also affect the third significant figure of the kinetic energy. 
Our measure of the radial distribution function does not feel appreciably
any of these effects, temperature, quantum statistics, or {\sl geometrical} 
gravity.

We then imagined a situation where we increase the mass of the star 
keeping all other parameters constant so to let the event horizon 
come closer to the patch and we noticed a diminution of both 
the kinetic and the potential energy of the plasma. Both the geometrical 
and the Coulomb potential energies diminish.

Comparing with the phase diagram of jellium in Euclidean space we see 
how in our case A
we are on the lower left corner of figure 1 of Ref. \cite{Brown2013} so 
that in order to see a crystallization we should at the same time wait 
for the star to cool down and observe the outer more rarefied part of the 
atmosphere.  


\section*{Author declarations}

\subsection*{Conflicts of interest}
None declared.

\subsection*{Data availability}
The data that support the findings of this study are available from the 
corresponding author upon reasonable request.

\subsection*{Funding}
None declared.

\bibliography{wdpimc}

\end{document}